# A Novel Real-Time Energy Management Strategy for Grid-Supporting Microgrid: Enabling Flexible Trading Power

Cunzhi Zhao
*Student Member, IEEE*
Department of Electrical and Computer Engineering
University of Houston
Houston, TX, USA
czhao20@uh.edu

Xingpeng Li
*Member, IEEE*
Department of Electrical and Computer Engineering
University of Houston
Houston, TX, USA
xli82@uh.edu

*Abstract*—In recent years, there has been significant growth of distributed energy resources (DERs) penetration in the power grid. The stochastic and intermittent features of variable DERs such as roof top photovoltaic (PV) bring substantial uncertainties to the grid on the consumer end and weaken the grid reliability. In addition, the fact that numerous DERs are widespread in the grid makes it hard to monitor and manage DERs. To address this challenge, this paper proposes a novel real-time grid-supporting energy management (GSEM) strategy for grid-supporting microgrid (MG). This strategy can not only properly manage DERs in a MG but also enable DERs to provide grid services, which enables a MG to be grid-supporting via flexible trading power. The proposed GSEM strategy is based on a 2-step optimization which includes a routine economic dispatch (ED) step and an acceptable trading power range determination step. Numerical simulations demonstrate the performance of the proposed GSEM strategy which enables the grid operator to have a dispatch choice of trading power with MG and enhance the reliability and resilience of the main grid.

*Index Terms*— Ancillary service, Battery energy storage system, Controllable Microgrid, Demand response, Distributed energy resources, Grid-supporting Microgrid, MG energy management, Model predictive control.

## Nomenclature

| | |
|---|---|
| $S_T$ | Set of Time period. |
| $S_G$ | Set of diesel generators. |
| $S_S$ | Set of energy storage systems. |
| $S_{WT}$ | Set of Wind Turbine. |
| $S_{PV}$ | Set of PV systems. |
| $c_{Gi}$ | Linear cost for diesel generator *i*. |
| $c_{Gi}^{NL}$ | No load cost for diesel generator *i*. |
| $c_{Gi}^{SU}$ | Start-up cost for diesel generator *i*. |
| $\Delta T$ | Size of a single dispatch interval. |
| $R_{percent}$ | Percentage of the backup power to the total power. |
| $E_{Si}^{Max}$ | Maximum energy capacity of ESS *i*. |
| $E_{Si}^{min}$ | Minimum energy capacity of ESS *i*. |
| $U_{Buyfgrid}^{t}$ | Status of purchasing power from main grid in time period *t*. |
| $U_{Selltgrid}^{t}$ | Status of selling power to main grid status in time period *t*. |
| $c_{Buyfgrid}^{t}$ | Wholesale electricity purchase price in time period *t*. |
| $c_{Selltgrid}^{t}$ | Wholesale electricity sell price in time period *t*. |
| $\rho$ | Air density. |
| $A$ | Area swept by the rotor blades. |
| $C_p$ | Power coefficient of the rotor. |
| $V$ | Velocity of the air. |
| $P_{Gi}^{Max}$ | Maximum capacity of generator *i*. |
| $P_{Gi}^{Min}$ | Minimum capacity of generator *i*. |
| $P_{Gi}^{t}$ | Output of generator *i* in time period *t*. |
| $U_{Chari}^{t}$ | Charging status of energy storage system *i* determined in economic dispatch in time period *t*. It is 1 if charging status; otherwise 0. |
| $U_{Disci}^{t}$ | Discharging status of energy storage system *i* determined in economic dispatch in time period *t*. It is 1 if discharging status; otherwise 0. |
| $U_{Gi}^{t}$ | Status of diesel generator *i* determined in economic dispatch in time period *t*. It is 1 if on status; otherwise 0. |
| $V_{Gi}^{t}$ | Start/Stop Status of diesel generator *i* determined in economic dispatch in time period *t*. It is 1 if start status; otherwise 0. |
| $P_{Buyfgrid}^{t}$ | Purchase from main grid power in time period *t*. |
| $P_{Selltgrid}^{t}$ | Sell to main grid power in time period *t*. |
| $P_L^t$ | Internal demand of MG at time period *t*. |
| $P_{DiscSi}^{t}$ | Discharging power of energy storage system *i* in time period *t*. |
| $P_{CharSi}^{t}$ | Charging power of energy storage system *i* in time period *t*. |
| $P_{Grid}^{Max}$ | Maximum thermal limit of tie-line between main grid and MG. |
| $P_{GiTari}^{t}$ | Target output for diesel generator *i* determined in economic dispatch at timer period *t*. |
| $P_{SiTari}^{t}$ | Target output for ESS *i* determined in economic dispatch at timer period *t*. |
| $P_{WTiTari}^{t}$ | Target output for WT *i* determined in economic dispatch at timer period *t*. |
| $P_{Gi}^{Ramp}$ | Ramping limit of diesel generator *i*. |
| $P_{Si}^{Max}$ | Maximum charge/discharge power of ESS *i*. |
| $P_{Si}^{Min}$ | Minimum charge/discharge power of ESS *i*. |
| $\eta_{Si}^{Disc}$ | Battery discharge efficiency of ESS *i*. |
| $\eta_{Si}^{Char}$ | Battery charge efficiency of ESS *i*. |
| $SOC_{Si}^{t}$ | State of charge of ESS *i* at time period *t*. |
| $OCslack$ | Overcharge slack variable. |
| $ODslack$ | Overdischarge slack variable. |
| $SOC_{low}$ | Minimum SOC level before overdischarge. |
| $SOC_{high}$ | Maximum SOC level before overcharge. |
| $\alpha_{Gi}$ | Adjusting parameter for diesel engine *i*. |
| $\alpha_{Si}$ | Adjusting parameter for ESS *i*. |
| $\alpha_{WTi}$ | Adjusting parameter for WT *i*. |

## I. Introduction

Microgrid (MG) is a local power system that connects distributed energy resources (DERs) and loads. MG are connected to main grid and have the capability to operate in either grid-connected mode or autonomously in isolated mode without the main grid. The development of smart control mechanisms for MG and DERs in recent decades has enlarged possibility that MG can enable flexible electricity consumption and provide ancillary services to the main grid [1]. Hence, a MG is considered to be grid-supporting if they operate in a manner that supports power grid reliability.

The DERs that can be integrated in MG includes but not limited to, traditional distributed energy resource such as diesel generator (DG), micro turbine (MT), fuel cell (FC) and renewable energy sources (RESs) such as photovoltaic (PV), wind turbine (WT) [2]. RESs have gained significant importance in today's world due to its environmentally friendly nature and the developed control technologies. However, it may lead to the unprecedented uncertainty to the main grid when the grid has high penetration of distributed RESs.

Energy management strategy of MG has been well studied in the literature. MG ancillary services provided to the utility grid are discussed in [3], a MG model is designed to provide the ancillary service (frequency control, spinning reserve, and non-spinning reserve) to the grid. In [4], a MG model is built to provide optimal scheduling to flatten the duck curve. Other technologies such as battery energy storage system (BESS) plays an importance role to mitigate the net-load fluctuation in MG as discussed in [5], [6], [7]. A model predictive control (MPC) also called rolling horizon method is applied to solve the mixed-integer linear programming (MILP) in [8], [9]. Although several MG energy management strategies are proposed in [3]-[9], they either focus on providing spinning reserve ancillary services to the grid or mitigating the internal fluctuation of MG. In addition, the proposed strategy in those papers cannot provide grid-supporting services that can assist main grid real-time load fluctuation or loss of generation especially with rapid growth of RESs. This could weaken the system reliability and lead to a substantial blackout.

To address the gap mentioned above, a real-time two-step grid-supporting energy management (GSEM) strategy is proposed for grid-connected MG in this paper. The proposed strategy can provide flexible grid-microgrid exchange power (trading power) to the grid. The proposed strategy consists of two steps: a routine economic dispatch (ED) optimization step along with an acceptable range of trading power determination (ARTPD) step. In the ED step, MPC method is employed to determine the targeted solutions which includes the outputs of adjustable energy resources (AERs) and the target trading power. In the ARTPD step, the exchange power maximization and minimization problems are formed with additional AERs' constraints based on targeted solutions obtained in the first step. Then, the upper bound and lower bound of the acceptable trading power range are determined by solving those maximization and minimization problems. As a result, the grid operator will receive a data package, from the grid-supporting MG, including a target trading power and a range of acceptable trading power with the associated cost deviating from the target trading power, which provides additional flexibility to grid operations.

The rest of this paper includes 4 sections. The proposed system model is presented in section II. Section III briefly discusses the two steps that determine the target trading power and the range of acceptable grid-microgrid trading power. Section IV includes case studies and discussions. Section V concludes the paper.

## II. SYSTEM MODELING

In this section, the system model is described for developing the proposed strategy. The DERs are classified as two categories; AER (controllable generation) and non-AER (generation is not controllable). Load modeling is included with residential load in a 24-hour time frame.

### A. Adjustable Energy Resource (AER)

Presently, DG are widely implemented as a backup energy resource in MG. DG is a traditional DER which is portable and provide fast response. DG has the capability to turn on automatically when the battery and RESs fail to maintain the load demand. DG is modelled with minimum and maximum physical limits in (1), ramp-up capability in (2) and ramp-down capability in (3). $C_G^t$ is the cost of DG represented in (4).

$$P_{Gi}^{Min} \leq P_{Gi}^t \leq P_{Gi}^{Max} \quad (i \in S_G, t \in S_T) \quad (1)$$

$$P_{Gi}^{t+1} - P_{Gi}^t \leq \Delta T \cdot P_{Gi}^{Ramp} \quad (i \in S_G, t \in S_T) \quad (2)$$

$$P_{Gi}^t - P_{Gi}^{t+1} \leq \Delta T \cdot P_{Gi}^{Ramp} \quad (i \in S_G, t \in S_T) \quad (3)$$

$$C_G^t = P_{Gi}^t c_{Gi} + U_{Gi} c_{Gi}^{NL} + V_{Gi} c_{Gi}^{SU} \quad (i \in S_G, t \in S_T) \quad (4)$$

Wind turbine develops rapidly due to government incentives and development of wind farm energy management. WT is the device that can convert the mechanical power of wind to electric power. The output power of wind turbine $P_{WTi}^t$ is defined in equation (5). $C_p$ is determined by the upstream velocity $V$ and downstream velocity $V_0$. When $\frac{V_0}{V} = \frac{1}{3}$, $C_p$ will have a maximum value of 0.59 [10]. Blades of wind turbine can be adjusted to a proper angle to achieve the maximum power output at different wind speed. However, at a given wind speed, the output power can be flexible by adjusting the blades' angle, which enables WT to become a AER.

$$P_{WTi}^t = \frac{1}{2}\rho A V^3 C_p \quad (5)$$

There are many types of energy storage system including supercapacitor, compressed air energy storage, flywheels, electrochemical battery and BESS [6]. In this paper, a BESS is modeled. The charging and discharging constraints are shown in (6)-(7); (8) indicates that the BESS have 3 modes, charging, discharging or sleeping. The equations of energy balance and SOC level of BESS are shown in (9) and (10). The degradation cost of BESS model includes two parts: (i) charging/discharging power cost $c_{spower}$ and (ii) overcharging/overdischarging cost $c_{soc}$. The degradation cost equation of BESS is shown in (11). The charging/discharging power cost is shown in (12). Overcharging and overdischarging slack variables are introduced to the model in order to form a soft constraint for limiting overcharging/overdischarging (13-16).

$$U_{Chari}^t \cdot P_{Si}^{Min} \leq P_{CharSi}^t \leq U_{Chari}^t \cdot P_{Si}^{Max}$$
$$(i \in S_S, t \in S_T) \quad (6)$$

$$U_{Disci}^t \cdot P_{Si}^{Min} \leq P_{DiscSi}^t \leq U_{Disci}^t \cdot P_{Si}^{Max}$$
$$(i \in S_S, t \in S_T) \quad (7)$$

$$U_{Disci}^t + U_{Chari}^t \leq 1 \quad (i \in S_G, t \in S_T) \tag{8}$$
$$E_{Si}^t - E_{Si}^{t-1} + \Delta T \cdot (P_{Disci}^{t-1}/\eta_{Si}^{Disc} - P_{Chari}^{t-1}\eta_{Si}^{Char}) = 0$$
$$(i \in S_G, t \in S_T) \tag{9}$$
$$SOC_{Si}^t = E_{Si}^t/E_{Si}^{Max} \quad (i \in S_S, t \in S_T) \tag{10}$$
$$C_{bess}^t = C_{power}^t + C_{soc}^t \tag{11}$$
$$C_{power} = (P_{Disci}^t/\eta_{Si}^{Disc} + P_{Chari}^t\eta_{Si}^{Char}) * c_{spower} \tag{12}$$
$$0 \leq ODslack \leq SOC_{low} \tag{13}$$
$$0 \leq OCslack \leq 1 - SOC_{high} \tag{14}$$
$$SOC_{low} - ODslack \leq SOC_{Si}^t \leq SOC_{high} + OCslack \tag{15}$$
$$C_{soc}^t = (DisCslack + OCslack) * c_{soc} \tag{16}$$

For grid-connected MG, the power distribution line between the MG and the main grid is called tie-line, which is also known as point of common coupling. The power that transmits on the tie-line can be assumed as one of the energy resources respect to MG. However, MG interacts with the main grid through the tie-line and can either consume or produce electricity. In this paper, the grid-microgrid exchange power is called trading power. Equation (17) represents the status of consuming or providing power of MG. The cost of trading power $C_{grid}^t$ is shown in (18). Thermal limits of tie-line are shown in (19)-(20).

$$U_{Buyfgrid}^t + U_{Selltgrid}^t \leq 1 \quad (i \in S_G, t \in S_T) \tag{17}$$
$$C_{grid}^t = P_{Buyfgrid}^t c_{Buyfgrid}^t - P_{Selltgrid}^t c_{Selltgrid}^t \quad (t \in S_T) \tag{18}$$
$$0 \leq P_{Buyfgrid}^t \leq U_{Buyfgrid}^t \cdot P_{Grid}^{Max} \quad (t \in S_T) \tag{19}$$
$$0 \leq P_{Selltgrid}^t \leq U_{Selltgrid}^t \cdot P_{Grid}^{Max} \quad (t \in S_T) \tag{20}$$

### B. Non-Adjustable Energy Resource (Non-AER)

Solar energy is mainly referring to the radiant light and heat from sunshine. The PV system converts the energy from sunshine to the DC current. This is followed by a DC-AC inverter which feed the grid or the energy storage system. Normally, the PV solar farm is considered as an AER since the power output can be controlled by the inverter. However, it is considered as Non-AER in this paper since the PV model is roof-top solar panels of residential houses which are not controllable by MG operator. The output power of PV system depends on several factors shown in (21). Which is mainly depends on solar irradiance $I$. $P_f$ is the shade ratio to the sunshine. $A_s$ represents the area of PV panel. $\eta_{pv}$ represents the efficiency of PV system.

$$P_{pvi}^t = IP_f A_s \eta_{pv} \tag{21}$$

### C. Load Modeling

Load model includes resident load and commercial load. Residential load represents the electricity that residents consume at any time. Commercial load represents the electricity that been used commercially, such as restaurant, mall & industry. In this MG model, only residential load is modeled as a 24-hour forecasting load with one-hour time span.

## III. FLEXIBLE TIE-LINE EXCHANGE POWER

As is discussed in [7], a constant trading power in a certain time period is proposed to reduce the tie-line power fluctuation to the main grid. In order to make the MG become grid-supporting, a real-time GSEM strategy is proposed to provide an acceptable trading power to the grid. The grid operator will have the option to select the dispatch point based on the provided acceptable trading power range and make the best choice with respect to the main grid.

### Step A. ED Optimization

The first step is to obtain the targeted solutions in the economic dispatch time period. MPC method is applied to optimize and solve a multi–interval MG economic dispatch MILP. The objective of the problem is to minimize the operation cost of the multi-interval but only the first interval's optimal solution would be implemented. The objective function is shown below:

$$f(cost) = \sum_{t \in S_T} C_G^t + C_{grid}^t + C_{bess}^t \tag{22}$$

where $C_G^t$, $C_{grid}^t$, $C_{bess}^t$ are introduced in Section II. Besides the constraints (1)-(21), there are two more constraints for the economic dispatch optimization problem:

$$P_{Buyfgrid}^t + \sum_{i \in S_G} P_{Gi}^t + \sum_{i \in S_{WT}} P_{WTi}^t + \sum_{i \in S_{PV}} P_{PVi}^t + \sum_{i \in S_S} P_{Disci}^t = P_{Selltgrid}^t + P_L^t + \sum_{i \in S_S} P_{Chari}^t \tag{23}$$

$$P_{Grid}^{Max} - P_{Buyfgrid}^t + P_{Selltgrid}^t + \sum_{G \in S_G}(P_{Gi}^{Max} - P_{Gi}^t) \geq R_{percent} P_L^t \quad (t \in S_T) \tag{24}$$

where equation (23) is the power balance equation of the MG. It is also employed in Step B1 and Step B2. Equation (24) is the constraint that forces MG to prepare sufficient backup power for the worst scenario.

### Step B1 Upper Bound of Acceptable Trading Power

This step is to determine the upper bound of the trading power that MG can address. The objective in equation (25) is to maximize the acceptable trading power, which can be also called as the upper bound of the trading power.

$$Max \; P_{Buyfgrid}^t - P_{Selltgrid}^t \tag{25}$$

The constraints of AERs are listed below:

$$P_{Gi}^t - P_{GiTari}^t \leq \alpha_{Gi} \cdot P_{Gi}^{Max} \quad (i \in S_G, t \in S_T) \tag{26}$$
$$P_{GiTari}^t - P_{Gi}^t \leq \alpha_{Gi} \cdot P_{Gi}^{Max} \quad (i \in S_G, t \in S_T) \tag{27}$$
$$(P_{Chari}^t - P_{SiTari}^t)t \leq \alpha_{si} \cdot E_{Si}^{Max} \quad (i \in S_S, t \in S_T) \tag{28}$$
$$(P_{SiTari}^t - P_{Chari}^t)t \leq \alpha_{si} \cdot E_{Si}^{Max} \quad (i \in S_S, t \in S_T) \tag{29}$$
$$(P_{Disci}^t - P_{SiTari}^t)t \leq \alpha_{si} \cdot E_{Si}^{Max} \quad (i \in S_S, t \in S_T) \tag{30}$$
$$(P_{SiTari}^t - P_{Disci}^t)t \leq \alpha_{si} \cdot E_{Si}^{Max} \quad (i \in S_S, t \in S_T) \tag{31}$$
$$P_{WTiTari}^t - P_{WTi}^{Max} \cdot \alpha_{wTi} \leq P_{WTi}^t \leq P_{WTiOpti}^t$$
$$(i \in S_W, t \in S_T) \tag{32}$$

where $\alpha_n$ is the preset parameter for each AER such as DE, BESS and WT. The targeted solutions of these AERs are solved in ED optimization. Equation (26)-(27) are the release constraints of DG. Therefore, DG can operate in an acceptable power range upon adjustable parameter $\alpha_{Gi}$ and target output of DG. Similar to DG, BESS can operate in an acceptable power range that determined by $\alpha_{si}$ and target BESS output (28)-(31). In equation (32), the acceptable power range of wind turbine is determined by $\alpha_{wTi}$ and target output of WT. $P_{WTiOpti}^t$ is the scheduled output power of WT. Note that all the constrains in previous ED optimization model are included in this step. The solution includes maximum value of tie-line trading power and the associated cost.

### Step B2 Lower Bound of acceptable Trading Power

Minimize optimization problem is formed up to get the acceptable lower bound of trading power. The constraints and

targeted outputs in step B2 are the same with Step B1. The objective function is shown in (33)

$$Min\ P_{Buyfgrid}^t - P_{Selltgrid}^t \quad (33)$$

which is to minimize the trading power. The solution of this problem includes the minimum value of tie-line trading power and the associated cost deviating from the target trading power.

Data Package: target trading power and an acceptable range of trading power with their associated cost are packed as a data array and sent to the grid. The grid operators will make the decision upon the dispatch point they need based on the provided data package. When the grid is suffering emergency load curtailment or loss of generation, it is a backup plan for grid operators to have an optional dispatch points.

## IV. CASE STUDIES

A typical grid-connected MG system is modeled in this paper as a testbed for the proposed strategy. The AERs include a 180kW DG (parameters shown in Table I), four 200kW WTs, and a 500kWh lithium-ion battery energy storage system. The battery charging/discharging efficiency is 90%. The non-AER includes 400 residential houses that have solar panel installed (5kW capacity per house). The load data represents 1,000 residential houses which are provided by Pecan Street Dataport [11] for a time period of 24 hours. The electricity price data are obtained from ECROT [12]. The price of selling electricity to the grid is set to 80% of the price purchasing electricity from the grid.

Table I. Diesel Engine Parameters

| Min Output | Max Output | Ramping | $c_G$ | NL Cost | SU Cost |
|---|---|---|---|---|---|
| 18 kW | 180 kW | 240 kW/h | 0.1 $/kWh | $ 3.4 | $ 5 |

In this work, python package Pyomo [13] and solver Gurobi [14] are used to specify and solve the economic dispatch optimization and acceptable range of trading power problems. Two scenarios that consist of both trading events between MG and main grid are under consideration in this paper.

*Scenario A: Time Period 15:15-15:30 Selling Power*

The target tie-line trading power for 15:15-15:30 is selling electricity to grid at a rate of 242.33 kW. BESS is on charging status with a power of 75 kW. Numerical tests are conducted on ten different combinations of adjusting parameter $\alpha_n$. Table II shows the test results for scenario A. The results associate with $\alpha_{Gi}\alpha_{Si}\alpha_{WTi} = 0$ serve as a benchmark. The power range represents the interval between lower bound and upper bound of the acceptable range of trading power (kW) while cost range ($/15min) represents the interval between their associated costs. It can be observed that as $\alpha_{Gi}, \alpha_{Si}, \alpha_{WTi}$ increase, the power range expands as well as the cost range. After comparing the result of test #1 and test #2, we can tell that after increasing the BESS adjusting parameter, the upper bound of the power range keeps the same while the lower bound of the power range decreases. Since the output solved by ED optimization is 75kW which is the maximum output of BESS, the charging rate is not able to gain higher even $\alpha_{Si}$ is increased. Similarly, after comparing the results of test #3 and test #4, we can tell that the lower bound of the power range keeps the same while the higher bound increases. The reason is that the targeted output of DG in this scenario reaches its maximum generation. However, DG can decrease the generation in order to expand the acceptable power range.

Table II. Results of 10 combinations of adjustable parameters for Scenario A

| No. | $\alpha_g$ | $\alpha_s$ | $\alpha_{wT}$ | Power Range | Cost Range |
|---|---|---|---|---|---|
| 0 | 0 | 0 | 0 | -242.33 | -1.82 |
| 1 | 0.05 | 0.01 | 0.05 | (-247.33, -193.33) | (-1.98, -0.79) |
| 2 | 0.05 | 0.02 | 0.05 | (-252.33, -193.33) | (-2.135, -0.79) |
| 3 | 0.05 | 0.02 | 0.08 | (-252.33, -169.33) | (-2.135, -0.17) |
| 4 | 0.08 | 0.02 | 0.08 | (-252.33, -163.93) | (-2.135, -0.164) |
| 5 | 0.08 | 0.05 | 0.1 | (-267.33, -147.93) | (-2.60, 0.245) |
| 6 | 0.1 | 0.05 | 0.1 | (-267.33, -144.33) | (-2.60, 0.251) |
| 7 | 0.1 | 0.08 | 0.1 | (-282.33, -144.33) | (-3.06, 0.251) |
| 8 | 0.12 | 0.08 | 0.1 | (-282.33, -140.73) | (-3.06, 0.254) |
| 9 | 0.15 | 0.08 | 0.1 | (-282.33, -135.33) | (-3.06, 0.258) |
| 10 | 0.15 | 0.1 | 0.1 | (-292.33, -135.33) | (-3.367, 0.258) |

*Scenario B: Time Period 19:45-20:00 Purchasing Power*

The target tie-line trading power for 19:45-20:00 is purchasing electricity from grid at a rate of 814.33 kW. BESS is on discharging status with a power of 20 kW. Table III. Shows the test results of scenario B. Similar to scenario A, while the adjusting parameter $\alpha_{Gi}, \alpha_{Si}, \alpha_{WTi}$ increasing, both power range and cost range expands. From test #1 and test #2 we can tell that the upper bound of the power range does not change, this is due to the minimum discharge power limit of BESS. We can observe the similar result when comparing test #3 and test #4 because the DG is in maximum generation. It can be concluded that the targeted solutions of AERs has a significant effect on the performance of the proposed model. If any output of AERs is on its maximum or minimum, then the acceptable power range can only expand in one direction.

Table III. Results of 10 combinations of adjustable parameters for Scenario B

| No. | $\alpha_g$ | $\alpha_s$ | $\alpha_{wT}$ | Power Range | Cost Range |
|---|---|---|---|---|---|
| 0 | 0 | 0 | 0 | 814.33 | 28.58 |
| 1 | 0.05 | 0.01 | 0.05 | (809.33, 863.33) | (28.45, 29.82) |
| 2 | 0.05 | 0.02 | 0.05 | (804.33, 863.33) | (28.33, 29.82) |
| 3 | 0.05 | 0.02 | 0.08 | (804.33, 887.33) | (28.33, 30.54) |
| 4 | 0.08 | 0.02 | 0.08 | (804.33, 892.73) | (28.33, 30.57) |
| 5 | 0.08 | 0.05 | 0.1 | (789.33, 908.73) | (27.95, 31.05) |
| 6 | 0.1 | 0.05 | 0.1 | (789.33, 912.33) | (27.95, 31.07) |
| 7 | 0.1 | 0.08 | 0.1 | (774.33, 912.33) | (27.58, 31.07) |
| 8 | 0.12 | 0.08 | 0.1 | (774.33, 915.93) | (27.58, 31.09) |
| 9 | 0.15 | 0.08 | 0.1 | (774.33, 921.33) | (27.58, 31.11) |
| 10 | 0.15 | 0.1 | 0.1 | (764.33, 921.33) | (27.33, 31.11) |

Further analysis based on the results of both scenarios are presented in Table IV, Table V, Figure 1, and Figure 2. Comparing Table IV with Table V, it can be overserved that when adjustable parameter $\alpha_{Gi}\alpha_{Si}\alpha_{WTi}$ are the same, the size of power range is the same. However, it may not be true for other scenarios. Equation for range efficiency $\eta_r$ is shown in (34), it means the average size of power range per dollar. Higher value of $\eta_r$ means each dollar can achieve higher size of power range, which is preferred. From Figure 1 and 2, we can tell that the range efficiency for scenario A fluctuates more than scenario B. Range efficiency $\eta_r$ in Table IV and V can be a standard to evaluate the efficiency that the model can provide acceptable power range. Figure 1 and Figure 2 provide an overview of the power range and the associated cost range. Bulk grid operator can use this information to determine whether it would deploy the grid-supporting service provided by MG. The driven force for MG to participate the proposed strategy is that MG can obtain more compensation/revenue from the main grid as it provides such a trading strategy to the main grid. If such services are not deployed by the bulk grid operators, MG can still operate at the optimal dispatch points which will minimize the MG operating cost. The pricing/compensation model needs to be further discussed, which is left for future work.

$$\eta_r = \frac{Size\ of\ Power\ Range}{Size\ of\ Cost\ Range} \quad (34)$$

Table IV. Analysis of Range Size for Scenario A

| Test Number | Size of Power Range | Size of Cost Range | Range Efficiency $\eta_r$ |
|---|---|---|---|
| 1 | 54 | 1.19 | 45.37815 |
| 2 | 59 | 1.345 | 43.86617 |
| 3 | 83 | 1.965 | 42.23919 |
| 4 | 88.4 | 1.971 | 44.85033 |
| 5 | 119.4 | 2.845 | 41.96837 |
| 6 | 123 | 2.851 | 43.14276 |
| 7 | 138 | 3.311 | 41.67925 |
| 8 | 141.6 | 3.314 | 42.72782 |
| 9 | 147 | 3.318 | 44.3038 |
| 10 | 157 | 3.625 | 43.31 |

Table V. Analysis of Range Size for Scenario B

| Test Number | Size of Power Range | Size of Cost Range | Range efficiency $\eta_r$ |
|---|---|---|---|
| 1 | 54 | 1.37 | 39.41606 |
| 2 | 59 | 1.49 | 39.59732 |
| 3 | 83 | 2.21 | 37.55656 |
| 4 | 88.4 | 2.24 | 39.46429 |
| 5 | 119.4 | 3.1 | 38.51613 |
| 6 | 123 | 3.12 | 39.42308 |
| 7 | 138 | 3.49 | 39.54155 |
| 8 | 141.6 | 3.51 | 40.34188 |
| 9 | 147 | 3.53 | 41.64306 |
| 10 | 157 | 3.78 | 41.53439 |

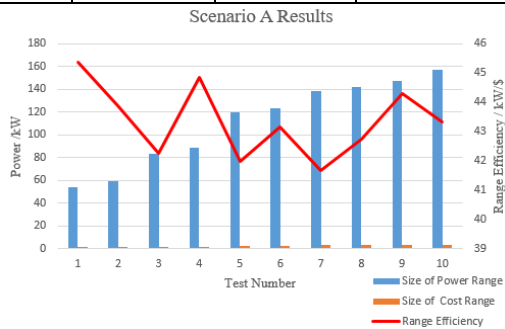

Figure 1. Data Analysis for Scenario A

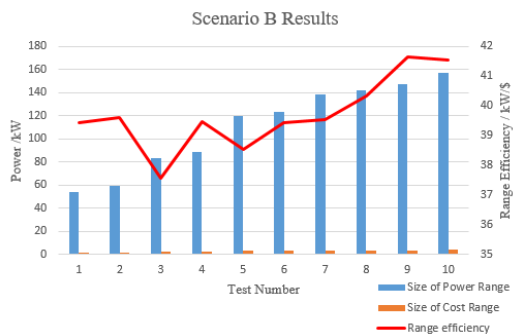

Figure 2. Data Analysis for Scenario B

## V. CONCLUSION

A novel real-time GSEM strategy for grid-supporting MG is proposed in this paper. The main contribution of this paper is that a novel energy management strategy is proposed to provide and quantify the grid-supporting service for the main grid. It consists of an ED optimization step and an acceptable range of trading power determination step. In ED optimization, targeted outputs of DERs and trading power are solved. In the acceptable range of trading power determination step, the upper bound and lower bound of acceptable power range are determined. Discussions about the relationship between acceptable power range and cost are covered as well. Numerical tests on AERs' parameters ensure that GSEM strategy can provide an acceptable range of trading power to the grid. Thus, the reliability of the grid can be enhanced by this grid service provided by MG. In summary, the proposed GSEM strategy further reinforces the grid's ability to handle the disturbance and enable the MG to be grid-supporting.

## VI. REFERENCES

[1] W. Shi, X. Xie, C. C. Chu and R. Gadh, "Distributed Optimal Energy Management in," *IEEE Transactions on Smart Grid,* vol. 6, no. 3, pp. 1137-1146, May 2015.

[2] W. Shi, N. Li, C. C. Chu and R. Gadh, "Real-Time Energy Management in MGs," *IEEE TRANSACTIONS ON SMART GRID,* vol. 8, no. 1, pp. 228-238, Jan 2017.

[3] A. F. Penaranda, P. E. Mosquera and S. F. Contreras, "Planning Model of MGs for the Supply of Ancillary Services to the Utility Grid," in *IEEE Milan PowerTech*, Milan, Italy, Jun 2019.

[4] A. Majzoobi and A. Khodaei, "Application of MGs in Supporting Distribution Grid Flexibility," *IEEE TRANSACTIONS ON POWER SYSTEMS,* vol. 32, no. 5, pp. 3660-3669, Sep 2017.

[5] H. Kanchev, D. Liu, F. Colas, V. Lazarov and B. Francois, *Energy Management and Operational Planning of a MG With a PV-Based Active Generator for Smart Grid Applications,* vol. 58, no. 10, pp. 4583-4592, 2011.

[6] M. T. Lawder, B. Suthar, P. W. Northrop, S. De, M. Hoff, O. Leitermann, M. L. Crow, S. Santhanagopalan and V. R. Subramanian, "Battery Energy Storage System (BESS) and Battery Management System (BMS) for Grid-Scale Applications," *Proceedings of the IEEE,* vol. 102, no. 6, pp. 1014-1030, Jun 2014.

[7] X. Li, G. Geng and Q. Jiang, "A hierarchical energy management strategy for grid-connected MG," in *IEEE PES General Meeting*, National Harbor, MD, July, 2014.

[8] A. Parisio, E. Rikos and L. Glielmo, "A Model Predictive Control Approach to MG Operation Optimization," *IEEE TRANSACTIONS ON CONTROL SYSTEMS TECHNOLOGY,* vol. 22, no. 5, pp. 1813-1827, 2014.

[9] R. Palma-Behnke, C. Benavides, F. Lanas, B. Severino, L. Reyes, J. Llanos and D. Saez, "A MG Energy Management System Based on the Rolling Horizon Strategy," *IEEE TRANSACTIONS ON SMART GRID,* vol. 4, no. 2, pp. 996-1006, June 2013.

[10] A. Sudhamshu, M. C. Pandey, N. Sunil, N. Satish and V. Mugundhan, *Numerical Study of effect of pitch angle onperformance characteristics of a HAWT,* vol. 19, no. 1, pp. 632-641, Aug, 2015.

[11] "Dataport Resources," May 2019. [Online]. Available: https://dataport.pecanstreet.org/academic..

[12] "ERCOT, Electric Reliability Council of Texas," [Online]. Available: http://www.ercot.com/.

[13] "Pyomo, Python Software packages.," Available: [Online]. Available: http://www.pyomo.org/ .

[14] "Gurobi Optimization, Linear Programming Solver," [Online]. Available: https://www.gurobi.com/.